\documentstyle[preprint,aps]{revtex}

\tightenlines
\begin{document}

\draft
\preprint{}
\title{Dynamically Spontaneous Symmetry Breaking  and
Masses of Lightest Nonet Scalar Mesons as Composite Higgs Bosons}
\author{ Yuan-Ben Dai and Yue-Liang  Wu  }
\address{ Institute of Theoretical Physics, Chinese Academy of Sciences, \\
 P.O. Box 2735, Beijing 100080, China }
\date{}
\maketitle

\begin{abstract}
 Based on the (approximate) chiral symmetry of QCD Lagrangian and the bound state assumption of effective meson
 fields, a nonlinearly realized effective chiral Lagrangian for meson
 fields is obtained from integrating out the quark fields by using the new finite regularization method.
 As the new method preserves the symmetry principles of the original theory and meanwhile keeps the finite
 quadratic term given by a physically meaningful characteristic energy scale $M_c$, it then leads to
 a dynamically spontaneous symmetry breaking in the effective
 chiral field theory. The gap equations are
 obtained as the conditions of minimal effective potential in the effective theory.
 The instanton effects are included via the induced interactions discovered by 't Hooft and found to play an
 important role in obtaining the physical solutions for the gap equations. The lightest nonet scalar mesons
 ($\sigma$, $f_0$, $a_0$ and $\kappa$) appearing as the chiral partners of the nonet pseudoscalar mesons are
 found to be composite Higgs bosons with masses below the chiral symmetry breaking scale
  $\Lambda_{\chi} \sim 1.2$ GeV. In particular, the mass of the singlet scalar (or the $\sigma$) is found to be
 $m_{\sigma} \simeq 677$ MeV.
\end{abstract}
%\narrowtex
%\end{titlepage}
%\maketitle

\newpage

\section{Introduction}

 The strong interaction between quarks is described by the SU(3) gauge theory,
 which is known as the chromodynamics (QCD). The asymptotic
 behavior of strong interaction at high energy has been successfully characterized by
 perturbative QCD. Though QCD was motivated from the studies of low energy dynamics
 of hadrons, the low energy dynamics of QCD remains unsolved due to the
 nonperturbative effects of strong interactions. In general, hadrons are considered
 to be the bound states formed
 by the quarks and gluons through the nonperturbative QCD effects. For lightest pseudoscalar mesons,
 the success of current algebra\cite{CM} with PCAC \cite{PCAC} is mainly because
 it reflects the (approximate) chiral invariance of the QCD lagrangian.
 While the (approximate) chiral symmetry $U(3)_L\times U(3)_R$ is found to be strongly broken
 down due to nonperturbative QCD effects. Many efforts have been paid to the issues
 such as:  how the chiral symmetry is dynamically broken down\cite{NJL}, how the instanton plays the role as it
 represents a quantum topological solution of nonperturbative QCD\cite{INST}, whether the effective meson
 theory should be realized as a linear $\sigma$ model\cite{LSM} or a non-linear $\sigma$ model, whether the lowest lying
 $U(3)_V$ nonet scalar mesons corresponds to the chiral partners of the lowest lying
 nonet pseudoscalar mesons, whether the isospinor scalar mesons $K_0^{\ast} (1430)$ \cite{PDG}
 are the lowest lying isospinor scalar mesons or there should exist other lighter isospinor
 scalar mesons $\kappa_0$\cite{E7912} that constitute the lowest lying nonet scalar mesons
 together with the isovector scalar mesons $a_0(980)$, the isoscalar scalar meson $f_0(980)$
 and the singlet scalar meson $f_0(400-1200)$ (or the $\sigma$)\cite{E7911}.

 Theoretically, some phenomenological models have been constructed to investigate the scalar sector.
 In this paper, we shall adopt the new finite regularization method proposed recently in ref.\cite{YLW} to
 derive an effective chiral Lagrangian for scalar and pseudoscalar mesons
 based on the (approximate) chiral symmetry of QCD Lagrangian and the bound
 state assumption of effective meson fields with including the instanton
 induced interactions discovered by t'Hooft\cite{INST,INST2}. For the purpose of the present paper, we then pay
 attention to study the chiral symmetry breaking mechanism and to predict the masses and mixing for the lightest
 scalar mesons. The advantages of the new finite regularization method are that it allows us to obtain the effective chiral
 Lagrangian which preserves gauge and Lorentz as well as translational invariance and
 meanwhile keeps the physically meaningful finite quadratic term. As a consequence,
 it leads the resulting effective chiral field theory to have a dynamically spontaneous symmetry breaking mechanism.
 Specifically, the gap equations are obtained as the conditions of minimal effective potential
 after spontaneous symmetry breaking. The important point in the
 new finite regularization method is that there appear two intrinsic mass scales, i.e., the characteristic
 energy scale (CES) $M_c$ and the sliding energy scale (SES) $\mu_s$. Here the CES $M_c$ is the
 basic energy scale below which the effective field theory becomes meaningful as the low energy dynamics of
  nonperturbative QCD. The SES $\mu_s$ reflects the energy scale on which the interesting physics processes
 are concerned. Because of these interesting features in the new finite regularization method,
 it manifestly distinguishes from either the dimensional regularization which
 cannot lead to the right gap equations though it preserves gauge and Lorentz invariance, or the naive cutoff
 regularization which destroys the gauge invariance though it could lead to the required gap equations.
 It will explicitly be shown that the resulting effective chiral Lagrangian is a nonlinearly
 realized chiral model, and the (approximate) chiral symmetry $U(3)_L\times U(3)_R$ of QCD Lagrangian is broken down through a
 dynamically spontaneous symmetry breaking mechanism. Here the lightest $U(3)_V$ nonet
 scalar mesons appearing as the chiral partners of the nonet pseudoscalar mesons
 are the composite Higgs bosons.
 Of particular, the instanton effect is found to play an important
 role in the dynamically spontaneous breaking of chiral symmetry $U(3)_L\times
 U(3)_R$. As expected, it not only provides a natural explanation why the singlet
 pseudoscalar meson $\eta'$ is much heavier than the octet pseudoscalar mesons, but also leads to a reasonable
 prediction for the mass spectrum of nonet scalar mesons. Actually, it is the instanton effect that
 makes the singlet scalar meson (the $\sigma$) to be much lighter than the octet scalar mesons,
 which is in contrast to the
 nonet pseudoscalar meson sector. In addition,
 it also results in consistent predictions for the light quark masses and the mixing angles between the singlet and
 octet neutral scalar and pseudoscalar mesons.

 \section{Effective Chiral Lagrangian and Dynamically Spontaneous Symmetry Breaking}

 Let us begin with the QCD Lagrangian with only light quarks

 \begin{eqnarray}
 {\cal L}_{QCD} = \bar{q}\gamma^{\mu}(i\partial_{\mu} + g_s G_{\mu}^a T^a)q -\bar{q}M q
  - \frac{1}{2} tr G_{\mu\nu}G^{\mu\nu}
 \end{eqnarray}
 where $q =(u, d, s)$ denote three light quarks and the summation over color degrees of freedom is understood.
 $G_{\mu}^a$ are the gluon fields with SU(3) gauge symmetry and $g_s$ is the running coupling constant.
 $M$ is the light quark mass matrix $M = diag.(m_1, m_2, m_3) \equiv diag.(m_u, m_d, m_s)$.
 In the limit $m_i \rightarrow 0$ (i=1,2,3), the Lagrangian has global
 $U(3)_L \times U(3)_R$ symmetry. It is known that the chiral
 $U(1)_L\times U(1)_R$ symmetry is broken down to $U(1)_V$ symmetry due to the quantum $U(1)_A$
 anomaly of QCD, namely the so-called instanton effect.
 The instanton-induced interactions were found to have the
 following form\cite{INST,INST2}

 \begin{equation}
 {\cal L}^{inst} = \kappa_{inst} e^{i\theta_{inst}} \det (-\bar{q}_R q_L )
 + h.c.
 \end{equation}
 where $\kappa_{inst}$ is a constant and contains the factor
 $e^{-8\pi^2/g^2}$.  Obviously, this instanton term breaks the
 chiral symmetry $U(1)_A$.

   The basic assumption in our present consideration is that at the chiral symmetry breaking
   scale ($\sim 1$ GeV) the effective Lagrangian contains
   not only the quark fields but also the effective meson fields describing bound states
   of strong interactions of gluons and quarks. After integrating
   over the gluon field at high energy scales,
   the effective Lagrangian at low energy scale is expected to have the following general form when keeping
   only the lowest order nontrivial terms
   \begin{eqnarray}
 {\cal L}_{eff} (q, \bar{q}, \Phi) & = & \bar{q}\gamma^{\mu}i\partial_{\mu}q
 + \bar{q}_L\gamma_{\mu}{\cal A}_L^{\mu}q_L +
 \bar{q}_R\gamma_{\mu}{\cal A}_R^{\mu}q_R - [\  \bar{q}_L (\Phi- M) q_R  + h.c. \  ]  \nonumber \\
 & + & \mu_m^2 tr \left(\Phi M^{\dagger} + M \Phi^{\dagger} \right)
 - \mu_f^2 tr \Phi \Phi^{\dagger}  + \mu_{inst} \left( \det \Phi + h.c. \right)
 \end{eqnarray}
 where $\Phi_{ij}$ are the effective meson fields which basically correspond to
 the composite operators $\bar{q}_{R j} q_{L i}$. ${\cal A}_L$ and ${\cal A}_R$ are introduced
 as the external source fields. The instanton induced interaction
 $\kappa_{inst} e^{i\theta_{inst}} \det (-\bar{q}_R q_L )$ has be mimicked
 by the term $ \mu_{inst} e^{i\theta_{inst}} \det \Phi $ \cite{INST2}.
  It is noticed that without considering
 the instanton-induced interaction term (or taking $\kappa_{inst} =0$), the integral over the
 effective meson fields $\Phi(x)$ leads to the four fermion interaction term
 and the resulting effective Lagrangian is then related to the Nambu-Jona-Lasinio
 model\cite{NJL}. Therefore without the instanton induced interactions,
 the above Lagrangian can be regarded as the bosonized
 Nambu-Jona-Lasinio model with the quark mass being given by
 \begin{eqnarray}
  \left( \frac{\mu_m^2}{\mu_f^2} -1 \right)\ M
 \end{eqnarray}
 it is clear that when taking $\mu_m^2 = 2
 \mu_f^2$, we then arrive at, after integrating over the effective meson field $\Phi_{ij}$,
 the well-known four quark interacting Nambu-Jona-Lasinio model with
 the well defined QCD current quark masses $M$.
 In this sense, the effective meson fields $\Phi_{ij}$ may be regarded as the auxiliary fields. Namely there
 is no kinetic term for the effective meson fields in our present
 considerations. Also the high order terms
 for the effective meson fields with dimension being equal and larger than four are not included assuming they are
 small and can be generated in loop diagrams. This indicates that only
 the lowest order nontrivial fermionic interaction terms are
 considered after integrating over the gluon field. For our purpose in the present paper,
  the vector and axial vector mesons are also not considered, only the scalar and pseudoscalar mesons are
  concerned. As the pseudoscalar mesons are known to be
  the would-be Goldstone bosons, the effective chiral field theory is naturally to be
  realized as a nonlinear model.
  Thus we may express the 18 effective meson fields of the $3\times 3$ complex matrix $\Phi(x)$ into the following form
  \begin{eqnarray}
  \Phi(x) & = & \xi_L(x) \phi(x) \xi_R^{\dagger}(x) , \quad  U = \xi_L(x)\xi_R^{\dagger}(x) = \xi^2_L(x) =
  e^{i\frac{2\Pi(x)}{f} }
   \nonumber \\
  & & \phi^{\dagger}(x) = \phi(x)= \sum_{a=0}^{a=9} \phi^{a}(x)T^a,
  \quad \Pi^{\dagger}(x) = \Pi(x)= \sum_{a=0}^{a=9} \Pi^{a}(x)T^a
  \end{eqnarray}
  where $T^a$ ($a=0, 1, \cdots, 8$) with $[T^a, T^b] = if^{abc} T^c$ and $2tr T^aT^b = \delta_{ab}$
  are the nine generators of $U(3)$ group. The fields $\Pi^a(x)$ represent the pseudoscalar mesons
   and $\phi^a(x)$ the corresponding scalar mesons.
  $f$ is a constant with mass dimension.

    The effective chiral Lagrangian for mesons is then obtained from integrating over the quark fields.
  The procedure for deriving the effective chiral Lagrangian of mesons can be formally expressed in terms of the
  generating functionals via the following relations
  \begin{equation}
   \frac{1}{Z} \int {\cal D}G_{\mu}{\cal D}q {\cal D}\bar{q}
  e^{i \int d^4 x {\cal L}_{QCD} } =  \frac{1}{\bar{Z}} \int {\cal D}\Phi {\cal D}q {\cal D}\bar{q}
  e^{i \int d^4 x  {\cal L}_{eff} (q, \bar{q}, \Phi) }  =
  \frac{1}{Z_{eff}} \int {\cal D} \Phi  e^{i \int d^4 x {\cal L}_{eff}(\Phi) }
  \end{equation}

   By applying the Schwinger's proper time technique\cite{JS} to the determinant of Dirac operator
 and using the new finite regularization method proposed in ref.\cite{YLW} for momentum integrals, we arrive at the
 following effective Lagrangian (a detailed derivation is presented in the appendix)
 \begin{eqnarray}
 {\cal L}_{eff}(\Phi)& = & \frac{1}{2} \frac{N_c}{16\pi^2}  tr T_0\ [\ D_{\mu}\hat{\Phi} D^{\mu}
 \hat{\Phi}^{\dagger} +
 D_{\mu}\hat{\Phi}^{\dagger} D^{\mu} \hat{\Phi} -(\hat{\Phi}\hat{\Phi}^{\dagger} - \bar{M}^2 )^2
 -(\hat{\Phi}^{\dagger}\hat{\Phi} - \bar{M}^2 )^2\ ] \nonumber  \\
 & + & \frac{N_c}{16\pi^2} M_c^2\ tr T_2\ [\ (\hat{\Phi}\hat{\Phi}^{\dagger} - \bar{M}^2 )
 + (\hat{\Phi}^{\dagger}\hat{\Phi} - \bar{M}^2 )
 \ ]
 \nonumber \\
 & + & \mu_m^2 tr \left(\Phi M^{\dagger} + M \Phi^{\dagger} \right)
 - \mu_f^2 tr \Phi \Phi^{\dagger}  + \mu_{inst} \left( \det \Phi + h.c. \right)
 \end{eqnarray}
 with $\hat{\Phi} = \Phi - M$, and  $\bar{M} = V - M = diag. (\bar{m}_1, \bar{m}_2 , \bar{m}_3 )$.
 Here  $\bar{m}_i =v_i -m_i$ is regarded as the dynamical quark masses, $v_i$ is supposed to be the vacuum
 expectation values (VEVs) of the scalar fields, i.e., $<\phi>= V = diag.(v_1, v_2, v_3)$ and
 to be determined from the conditions of minimal effective potential
 in the effective chiral Lagrangian ${\cal L}_{eff}(\Phi)$.
  Here we have only kept the lowest terms needed for our purpose
 in the present paper. Where the two diagonal matrices $T_0=diag.(T_0^{(1)},T_0^{(2)}, T_0^{(3)})$
 and $ T_2 = diag.(T_2^{(1)}, T_2^{(2)}, T_2^{(3)})$
 arise from the integral over the loop momentum by using the new finite regularization method. They are given by the
 following form
  \begin{eqnarray}
   T_0^{(i)}(\frac{\mu^2_i}{M_c^2}) & = & \ln \frac{M_c^2}{\mu^2_i } - \gamma_w + y_0(\frac{\mu^2_i}{M_c^2}) \\
     T_2^{(i)}(\frac{\mu^2_i}{M_c^2}) & = &  1 - \frac{\mu_i^2}{M_c^2} [ \ \ln \frac{M_c^2}{\mu^2_i}
     - \gamma_w + 1 +   y_2(\frac{\mu^2_i}{M_c^2}) \ ]
  \end{eqnarray}
   with
  \begin{eqnarray}
      & & y_0 (x) = \int_0^x d \sigma\ \frac{1 - e^{-\sigma} }{\sigma} \\
      & & y_1 (x) = \frac{1}{x} \left(e^{-x} - 1 + x\right) \\
      & & y_2 (x) = y_0 (x) - y_1(x)
  \end{eqnarray}

  Note that $M_c$ is the characteristic energy scale from which the nonperturbative QCD effects start to play an
  important role and the effective chiral field theory is considered to be valid below the scale $M_c$.
  We have also used the definitions
  \begin{equation}
  \mu_i^2 = \mu_s^2 + \bar{m}_i^2 , \qquad \bar{m}_i =v_i -m_i
  \end{equation}
 with $\mu_s^2$ the sliding energy scale arising from the new finite regularization, it is usually taken to be
 at the energy scale at which the physical processes take place. As it is easy to see that the equal
 mass case $m_u=m_d=m_s$ must lead to the equal VEVs $v_1=v_2=v_3$, one may then write the VEVs in terms of
 the following general form
\begin{equation}
v_i = v_o + \beta\ m_i  , \qquad i= 1,2,3 \quad \mbox{or} \quad i=u, d, s
\end{equation}

 Let us now focus on the effective potential which may be reexpressed as the following general form
\begin{eqnarray}
 V_{eff}(\Phi) & = &  - tr  \hat{\mu}_m^2 \left(\Phi M^{\dagger} + M \Phi^{\dagger} \right)
 + \frac{1}{2} tr \hat{\mu}_f^2 (\Phi\Phi^{\dagger} + \Phi^{\dagger}\Phi )  \nonumber \\
   & + & \frac{1}{2} tr \lambda [ \ (\hat{\Phi}\hat{\Phi}^{\dagger} )^2 + (\hat{\Phi}^{\dagger}\hat{\Phi} )^2\  ]
  -\mu_{inst} \left( \det \Phi + h.c. \right)
 \end{eqnarray}
with $\hat{\mu}_f^2 $, $\hat{\mu}_m^2$ and $\lambda$ the three diagonal matrices
\begin{eqnarray}
\hat{\mu}_f^2 &  =  & \mu_f^2 - \frac{N_c}{8\pi^2} \left( M_c^2 T_2 + \bar{M}^2 T_0 \right)  \\
\hat{\mu}_m^2 &  =  & \mu_m^2 - \frac{N_c}{8\pi^2} \left( M_c^2 T_2 + \bar{M}^2 T_0 \right)
 \\
\lambda & = &  \frac{N_c}{16\pi^2} T_0
\end{eqnarray}
Taking the nonlinear realization $\Phi(x) = \xi_L(x) \phi(x) \xi_R^{\dagger}(x)$ with supposing that the minimal
of the above effective potential occurs at the point $<\phi>= V = diag.(v_1, v_2, v_3)$, and writing the scalar
fields as
\begin{equation}
\phi(x) = V + \varphi(x)
\end{equation}
we then obtain three minimal conditions
\begin{eqnarray}
 -\left(\hat{\mu}_f^2\right)_i v_i + \left(\hat{\mu}_m^2\right)_i m_i - 2 \lambda_i \bar{m}_i^3 + \mu_{inst}
\bar{v}^3/v_i = 0, \quad i=1, 2, 3
\end{eqnarray}
  with $\bar{v}^3 = v_1 v_2 v_3$. For convenience of discussions,
  it is  useful to decompose the diagonal matrices
  $\mu^2$, $\hat{\mu}_f^2$, $\hat{\mu}_m^2$ and $\lambda$ into two parts that are
  independent of and dependent on the current quark masses $m_i$ ($i=u,d,s$). Practically, it can be done by
  making an expansion in terms of the current quark masses
   \begin{eqnarray}
   & & \mu_i^2 =  \mu_o^2  + 2(\beta - 1) v_o \tilde{m}_i, \qquad \mu_o^2 = \mu_s^2 + v_o^2 ,
   \quad \tilde{m}_i = m_i [ 1 + (\beta -1) m_i/(2v_o) ]  \\
   & & \left(\hat{\mu}_f^2\right)_i = \bar{\mu}_f^2 + 2\mu_{f o}\tilde{m}_i [\  1 + \sum_{k=1} \alpha_k
   \left( \frac{\tilde{m}_i}{\mu_o} \right)^k (\beta - 1)^k \  ]  \\
   & & \left(\hat{\mu}_m^2\right)_i  = \bar{\mu}_m^2  + 2\mu_{f o}\tilde{m}_i [\  1 + \sum_{k=1} \alpha_k
   \left( \frac{\tilde{m}_i}{\mu_o} \right)^k (\beta - 1)^k \  ] \\
  & & \lambda_i = \bar{\lambda} -  \lambda_o \sum_{k=1} \beta_k \left( \frac{\tilde{m}_i}{\mu_o} \right)^k (\beta -
  1)^k , \qquad \lambda_o =  \frac{N_c}{16\pi^2}
  \end{eqnarray}
  The unknown seven parameters are $\bar{\mu}_f^2$, $\bar{\mu}_m^2$, $\mu_{inst}$, $v_o$, $\beta$
  $\mu_s$ and $\bar{\lambda}$ in addition to the three current quark masses.
  Three of them are determined by the three minimal conditions. $\mu_{inst}$ is in principle calculable and
  will actually be fixed by the $\eta'$ mass. $v_o$ is related to the pion decay constant $f$.
  $\bar{\lambda}$ is given by the characteristic energy scale $M_c$ at which the effective chiral field theory
  become meaningful, namely $M_c$ is at the same order as the chiral symmetry breaking scale
  $M_c \sim 4\pi f$. The sliding energy scale $\mu_s$ runs to the scale at the order of dynamically spontaneous symmetry
  breaking scale $v_o$.

   Solving the three equations of the minimal conditions with keeping only the nonzero
   leading terms in the expansion of current quark masses, we then obtain
   the following three constraints from the minimal conditions
   \begin{eqnarray}
    & &  v_o (1 - \epsilon_o) (\beta - 1)^2/\beta^2 \simeq v_{inst}/3 \\
    & &  2\bar{\lambda}(v_{inst}v_3 - v_o^2 )  \simeq \bar{\mu}_{f}^2   \\
   & &  2 \beta \bar{\lambda} v_o^2 + 6(\beta -1) \bar{\lambda} v_o^2 +
    2\lambda_o (\beta -1) v_o^2 ( 1- \frac{2v_o^2}{\mu_o^2} ) \simeq \bar{\mu}_m^2  - 2\beta \bar{\mu}_{f}^2
   \end{eqnarray}
   where we have neglected the small quark masses $m_u$ and $m_d$ and introduced the definitions
   \begin{eqnarray}
   & &  \epsilon_o = \frac{\lambda_o}{\bar{\lambda}}  [\
    \left(\frac{2v_o^2}{\mu_o^2}-1 \right)\left( 1 -
    \frac{1}{3}\frac{v_o^2}{\mu_o^2} \right) - \frac{1}{3}
    \frac{2v_o}{\mu_o} \alpha_1 (1- r) + r + (\beta - 1)\left(\frac{2v_o^2}{\mu_o^2} -
    \frac{1}{3}\frac{\bar{\lambda}}{\lambda_o} \right) \frac{m_s}{v_o} \ ] \\
   & & \mu_{fo} \equiv 2\lambda_o v_o (\beta -1 ) ( 1 -r) \\
   & & \mu_{inst} \equiv 2 \bar{\lambda} v_{inst}
   \end{eqnarray}
  Here $r$ and $\alpha_1$ are given by
  \begin{eqnarray}
   & & r = \frac{\mu_s^2}{\mu_o^2} - \frac{\mu_o^2}{M_c^2} [ 1 +  \frac{\mu_s^2}{\mu_o^2}
   + O( \frac{\mu_o^2}{M_c^2}) ] \\
   & & \alpha_1 (1-r) = \frac{2v_o}{\mu_o}\ [ \ \frac{\mu_s^2}{2\mu_o^2} +  O( \frac{\mu_o^2}{M_c^2}) \  ]
   \end{eqnarray}
 Note that in obtaining eq.(25) one needs to keep terms to the order of $m_i^2$.
  The parameters $\bar{\mu}_{m}^2$ and $\bar{\mu}_f^2 $ are related to the initial parameters in the effective
 potential and the characteristic energy scale via the following relations
  \begin{eqnarray}
  \bar{\mu}_{f}^2 & = &  \mu_f^2 - \frac{N_c}{8\pi^2} \left( M_c^2 T_2(\frac{\mu_o^2}{M_c^2})
  + v_o^2 T_0 (\frac{\mu_o^2}{M_c^2})   \right)  \\
\bar{\mu}_m^2 &  =  & \mu_m^2 - \frac{N_c}{8\pi^2} \left( M_c^2 T_2(\frac{\mu_o^2}{M_c^2}) + v_o^2
T_0(\frac{\mu_o^2}{M_c^2}) \right) \\
 \bar{\lambda} & = & \frac{N_c}{16\pi^2} T_0(\frac{\mu_o^2}{M_c^2}), \qquad
 T_0(\frac{\mu_o^2}{M_c^2}) =  \ln \frac{M_c^2}{\mu^2_o } - \gamma_w + y_0(\frac{\mu^2_o}{M_c^2})
   \end{eqnarray}
 From the normalization of the kinetic term for the pseudoscalar mesons, we have the relation
   \begin{equation}
  \bar{\lambda} v_o^2 = f^2 /4
   \end{equation}

   Note that when ignoring the instanton effects, i.e., taking $v_{inst} = 0$, one sees that
   the minimal conditions lead to the usual gap equation
   \begin{eqnarray*}
   & & \frac{N_c}{8\pi^2  \mu_f^2}\ [ \ M_c^2 - \mu_o^2  \left( \ln \frac{M_c^2}{\mu^2_o } - \gamma_w + 1
   + y_2(\frac{\mu^2_o}{M_c^2}) \right ) \  ]
   = 1
   \end{eqnarray*}
   However, the anomalous large mass of $\eta'$ implies that the instanton effects are important.
   Therefore the conditions of  minimal effective potential must be modified after including the
   instanton effects . They may be regarded as the generalized gap equations
   and their effects will be discussed in the next section. So far, we have explicitly shown the
   mechanism of dynamically spontaneous symmetry breaking.

 \section{Masses of light quarks and Lightest Nonet Scalar Mesons}

   To make numerical predictions for the masses of scalar and pseudoscalar mesons, one needs to
   solve the generalized gap equations. For that one must have the knowledge for the three
   initial parameters $\mu_f^2$, $\mu_m^2$ and $\mu_{inst}$ appearing in the original effective
   Lagrangian. In addition, one should also know the intrinsic mass scales $M_c$ and $\mu_s$.
   In principle, they all should be calculable from QCD and depend on the QCD running scale $\mu$
   the basic QCD scale $\Lambda_{QCD}$ as well as the light quark masses $m_u$, $m_d$ and $m_s$.
   Practically, one may choose the parameters $v_o$,  $\beta$, $M_c$, $\mu_s$ and the light quark
   masses as input, this is because
   such a set of parameters are much directly related to the low energy phenomena.

    As we have shown in the previous section that to have well defined QCD current quark masses without
    considering the instanton interaction term, it requires that
    \begin{eqnarray}
   \left( \frac{\mu_m^2}{\mu_f^2} -1 \right)\ M = M, \quad \mbox{i.e.} \quad \mu_m^2 = 2 \mu_f^2
 \end{eqnarray}
   which reduces one parameter.

  Also when omitting the instanton interaction term, the
  auxiliary fields $\Phi_{ij}$ are found from the effective
  Lagrangian eq.(3) to be given by the quark fields as follows
   \begin{eqnarray}
    \Phi_{ij} =  -\frac{1}{\mu_f^2} \bar{q}_{R j} q_{L i} + \frac{\mu_m^2 }{\mu_f^2} M _{ij}
 \end{eqnarray}
 Assuming the quark condensate is almost flavor independent,
 i.e., $<\bar{u}u> \simeq <\bar{d}d> \simeq <\bar{s}s>$, we then obtain by combining
 the above condition $\mu_m^2 = 2 \mu_f^2$ the second condition
  \begin{eqnarray}
   \beta =  \frac{\mu_m^2 }{\mu_f^2}  = 2
  \end{eqnarray}
    so that the dynamical quark masses have the simple form
   \begin{eqnarray}
   \bar{m}_i= v_i - m_i = v_o + (\beta - 1) m_i =   v_o +  m_i ,\qquad  i =u, d, s
   \end{eqnarray}
    which may also be regarded as a kind of constituent quark masses after dynamically spontaneous symmetry
    breaking. Where $v_o$ is caused by the quark condensate
   \begin{eqnarray}
    v_o = - \frac{1}{2\mu_f^2} <\bar{q} q >, \qquad q =u, d, s
   \end{eqnarray}
   It is supposed that the inclusion of the instanton interaction term do not significantly change the above
   two conditions. Namely, we take $ \beta \simeq \mu_m^2/\mu_f^2 \simeq 2$ as a good approximation
   to reduce two parameters. This is because the instanton term is found to be much smaller than the quadratic term
   $\mu_{inst}/\mu_f \simeq 0.06$ (see below).

   To determine the remaining parameters, we consider the following constraints.
   Two constraints arise from the pseudoscalar sector. One is due to the normalization of the kinetic term
   (eq.(36))
  \begin{eqnarray}
  T_0  v_o^2 = \frac{(4\pi f )^2}{4N_c} \equiv \bar{\Lambda}_f^2 \simeq (340 \mbox{MeV})^2
  \end{eqnarray}
  The other is from the mass matrix for the isoscalar  and singlet pseudoscalar mesons $\eta_8$ and $\eta_0$ (see below),
  the trace of the mass matrix leads to
  \begin{eqnarray}
  v_{inst}v_3  = \frac{1}{6} (m_{\eta_8}^2 + m_{\eta_0}^2 - 2 m_K^2 ) = \frac{1}{6} (m_{\eta}^2 + m_{\eta'}^2 - 2 m_K^2 )
  \simeq (348 \mbox{MeV})^2
  \end{eqnarray}
  Where we have used $f \simeq 94$ MeV and the experimental data
  $m_K \simeq 496 $ MeV, $m_{\eta} \simeq 548$ MeV and $m_{\eta'} \simeq 958$ MeV.

  In the scalar sector,  the well measured scalar meson $a_0 (980)$ will provide a constraint to the VEV $v_o$
  with $v_o \sim \bar{\Lambda}_f = 340 $ MeV. This indicates that no large loop corrections arise from
  the logarithmic term in our present considerations, namely
 \begin{eqnarray}
  T_0(\mu_o^2/M_c^2) \simeq 1 \quad \mbox{i.e.} \quad \bar{\lambda} \simeq \lambda_o = N_c/(16 \pi^2)
  \end{eqnarray}

   With the above constraints together with the three minimal conditions, the parameters
   are determined to be
  \begin{eqnarray}
 & &  v_o \simeq \bar{\Lambda}_f \simeq 340 MeV  \nonumber \\
  & &  \mu_m^2 = 2\mu_f^2 \simeq (204 \mbox{MeV} )^2  \nonumber \\
 & &   M_c \simeq  922 MeV , \qquad \mu_s \simeq  333 MeV \nonumber \\
  & &   v_{inst} \simeq 210 MeV , \quad \mbox{or} \quad  \mu_{inst} = 2\bar{\lambda} v_{inst} \simeq 8.0 MeV \nonumber \\
  & & <\bar{q}q > = - (242 MeV)^3 , \qquad  m_s \simeq 117 MeV
  \end{eqnarray}
 Here the resulting quark condensation is consistent with the
 one from QCD.

  We are now in the position to make predictions on the masses and mixing for the scalar mesons, pseudoscalar mesons
  and/or light quark masses. To be manifest, let us first write down the scalar and pseudoscalar meson matrices
 \begin{equation}
 \sqrt{2}\varphi =\left(
\begin{array}{ccc}
\frac{a^0_0}{\sqrt{2}} +\frac{1}{\sqrt{6}}f_8 +\sqrt{\frac{1}{3}}f_s
& a_0^+ & \kappa_0^+  \\[2mm]
a_0^- &
- \frac{a_0^0}{\sqrt2} +\frac{1}{\sqrt{6}}f_8 +\sqrt{\frac{1}{3}}f_s &  \kappa_0^0 \\[2mm]
\kappa_0^- & \bar{\kappa}_0^0 &  -\frac{2}{\sqrt{6}}f_8 +\sqrt{\frac{1}{3}}f_s
\end{array} \right)\,,
\end{equation}
 and
  \begin{equation}
 \sqrt{2}\Pi=\left(
\begin{array}{ccc}
\frac{\pi^0}{\sqrt{2}} +\frac{1}{\sqrt{6}}\eta_8 +\sqrt{\frac{1}{3}}\eta_0
& \pi^+ & K^+  \\[2mm]
\pi^- &
- \frac{\pi^0}{\sqrt2} +\frac{1}{\sqrt{6}}\eta_8 +\sqrt{\frac{1}{3}}\eta_0 &  K^0 \\[2mm]
K^- & \bar{K}^0 &  -\frac{2}{\sqrt{6}}\eta_8 +\sqrt{\frac{1}{3}}\eta_0
\end{array} \right)\,,
\end{equation}

 Keeping to the leading order of current quark masses, we have
 \begin{eqnarray}
& &  m_{\pi^{\pm}}^2 \simeq \frac{2\mu_P^3}{f^2} (m_u + m_d ) \\
& &  m_{K^{\pm}}^2 \simeq \frac{2\mu_P^3}{f^2} (m_u + m_s ) \\
 & &  m_{K^{0}}^2 \simeq \frac{2\mu_P^3}{f^2} (m_d + m_s ) \\
 & &  m_{\eta_8 }^2 \simeq \frac{2\mu_P^3}{f^2} [ \  \frac{1}{3}( m_u + m_d ) + \frac{4}{3} m_s \  ]
 = \frac{1}{3} ( 4 m_K^2 - m_{\pi}^2 )  \\
  & &  m_{\eta_8\eta_0 }^2 \simeq -\frac{2\mu_P^3}{f^2} \frac{\sqrt{2}}{3} [ \ 2m_s -( m_u + m_d ) \  ]
 =  - \frac{2\sqrt{2}}{3} ( m_K^2 - m_{\pi}^2 )  \\
 & &  m_{\eta_0 }^2 \simeq \frac{2\mu_P^3}{f^2} \frac{2}{3}( m_u + m_d  + m_s  )
 +  \frac{12 \bar{v}^3}{f^2} \mu_{inst}
 = \frac{1}{3} ( 2 m_K^2 + m_{\pi}^2 ) +  \frac{24 \bar{v}^3}{f^2} \bar{\lambda} v_{inst}
 \end{eqnarray}
 where $\mu_P^3$ is given by
 \begin{eqnarray}
\mu_P^3 = (\bar{\mu}_m^2 + 2 \bar{\lambda} v_o^2 ) v_o \simeq 12
\bar{\lambda} v_o^3 \simeq  3 v_o f^2
 \end{eqnarray}
 By taking the numerical value for the relevant parameter $v_o \simeq 340 $ MeV and using the experimental
 data for the pion meson mass $m_{\pi} \simeq 139$ MeV and the mass square difference between the neutral and charged kaons
  $m_{K^0}^2 -m_{K^{\pm}}^2 \simeq (63 \mbox{MeV} )^2$, we then arrive at
  the following predictions
 \begin{eqnarray}
 & & m_u + m_d \equiv 2 \bar{m} \simeq 9.5\ \mbox{MeV}, \qquad  m_d - m_u \simeq 1.95 \mbox{MeV} \\
 & & m_u \simeq 3.8 \mbox{MeV}, \qquad m_d \simeq 5.7 \mbox{MeV}, \qquad m_s/m_d \simeq
 20.5
 \end{eqnarray}
  for the up and down quark masses, and
 \begin{eqnarray}
 & &  m_{K^0} \simeq 500 \mbox{MeV}, \qquad   m_{K^{\pm}} \simeq 496 \mbox{MeV}  \\
 & &  m_{\eta} \simeq 503 \mbox{MeV}, \qquad   m_{\eta'} \simeq 986 \mbox{MeV}
 \end{eqnarray}
 for the pseudoscalar meson masses, as well as
 \begin{eqnarray}
  \tan 2\theta_P = 2 \sqrt{2} [ 1 - \frac{9 v_{inst}v_3}{m_K^2 - m_{\pi^2}} ]^{-1} , \qquad \theta_P \simeq
  -18^{o}
 \end{eqnarray}
 for  the mixing angle $\theta_P$ between $\eta$ and $\eta'$
 mesons, which is defined as
 \begin{eqnarray}
 \eta_8 = \cos \theta_P \ \eta +   \sin \theta_P \ \eta'  \nonumber \\
 \eta_0 = \cos \theta_P \ \eta'  -  \sin \theta_P \ \eta
 \end{eqnarray}

  The masses and mixing of scalar mesons are given by
  \begin{eqnarray}
 & &  m_{a_0^{\pm}}^2 \simeq  m_{a_0^{0}}^2 \simeq 2 (2\bar{m}_u + \bar{m}_d ) \bar{m}_u  + 2v_{inst}v_3  \\
 & &  m_{k_0^{\pm}}^2 \simeq 2 (2\bar{m}_u + \bar{m}_s ) \bar{m}_u  + 2v_{inst}v_2  \\
 & &  m_{k_0^{0}}^2 \simeq 2 (2\bar{m}_d + \bar{m}_s ) \bar{m}_d  + 2v_{inst}v_1  \\
 & &  m_{f_8}^2 \simeq \bar{m}_u^2 + \bar{m}_d^2 + 4 \bar{m}_s^2   + \frac{2}{3} v_{inst}(2v_1 + 2v_2 -v_3 ) \\
 & &  m_{f_s}^2 \simeq 2 (\bar{m}_u^2 + \bar{m}_d^2 + \bar{m}_s^2)   - \frac{4}{3} v_{inst}( v_1 + v_2 + v_3 )  \\
 & &  m_{f_s f_8 }^2 \simeq \sqrt{2} ( 2\bar{m}_s^2 - \bar{m}_u^2  - \bar{m}_d^2 )
 - \frac{\sqrt{2}}{3} v_{inst} (2v_3 - v_1 - v_2 ) \\
 & & \tan 2\theta_S = \frac{2  m_{f_sf_8 }^2 }{ m_{f_s}^2 - m_{f_8}^2 }
  \end{eqnarray}
 Here the mixing angle $\theta_S$ is defined as
  \begin{eqnarray}
 f_8 = \cos \theta_S \ f_0 +   \sin \theta_S \ \sigma  \nonumber \\
 f_s = \cos \theta_S \ \sigma  -  \sin \theta_S \ f_0
 \end{eqnarray}
  where we have ignored the mixing between $a_0^0$ and $f_8$ ($f_0$) as it is proportional to $v_1 - v_2 \sim
  m_u -m_d $.  Inputting the values $\bar{m}_u \simeq \bar{m}_d = v_o + \bar{m} \simeq 345$ MeV,
  $\bar{m}_s= v_o + m_s \simeq 457 $ MeV, $v_{inst}v_3 \simeq (348 MeV)^2$,
  $v_1 \simeq v_2  \simeq v_o + 2\bar{m} \simeq 350$ MeV, $v_{inst} \simeq 210$ MeV, we arrive at
  the following numerical predictions for the  masses and mixing of the scalar mesons
  \begin{eqnarray}
 & &    m_{a_0} \simeq 978 \ \mbox{MeV},  \qquad   m_{a_0}^{exp.} = 984.8 \pm 1.4 \  \mbox{MeV}\  \cite{PDG} \\
 & &   m_{\kappa_0} \simeq 970\ \mbox{MeV},   \qquad
    m_{\kappa_0}^{exp.} = 797\pm 19\pm 43  \  \mbox{MeV}\  \cite{E7912} \\
 & &   m_{f_0} \simeq 1126\ \mbox{MeV},  \qquad   m_{f_0}^{epx.} = 980 \pm 10 \  \mbox{MeV}\  \cite{PDG}\\
 & &   m_{\sigma} \simeq 677\ \mbox{MeV}, \qquad  m_{\sigma}^{exp.} = (400 - 1200) \  \mbox{MeV}\  \cite{PDG} \\
 & &   \theta_S \simeq  - 18^{o}
  \end{eqnarray}

 The above predictions are at the leading order approximation in the expansion of current quark masses and
 also at the tree level in the effective chiral field theory. Now some important features become clear.
 In contrast to the singlet pseudoscalar meson $\eta'$ which is much heavier than the $\eta$ meson,
 the singlet scalar meson $\sigma$ is much lighter than $f_0$. Also the isospinor
 scalar meson $\kappa_0$ is below 1 GeV and likely
 lighter than the isovector scalar meson $a_0$. This feature has also been observed by many groups\cite{kappa}.
 All such features are mainly due to the instanton effects, which can easily be seen from
 their mass formulae eqs.(61-66). Note that the isoscalar meson mass
 $m_{f_0}$ is somehow larger than the experimental data by about $15\%$ at the leading order, it is of interest to
 investigate the contributions from possible higher order terms.

 \section{conclusions}

 Starting from the effective Lagrangian of chiral quarks with effective meson fields as
 bosonized auxiliary fields at the chiral symmetry breaking scale, which is assumed to
 be resulted from integrating out the gluon fields,
 a nonlinearly realized effective chiral Lagrangian for meson fields has been obtained
 from integrating over the quark fields by using the new finite regularization method.
 It has been shown that the resulting effective chiral Lagrangian can lead to a dynamically
 spontaneous symmetry breaking mechanism. This is because the new finite regularization
 method keeps the physically meaningful finite quadratic term and meanwhile
 preserves the symmetry principles of original theory.
 After the chiral symmetry $U(3)_L\times U(3)_R$ is spontaneously broken down, the effective chiral Lagrangian
 contains, in addition to the three current quark masses $m_u$, $m_d$ and $m_s$, four basic
 parameters, $v_o$, $\beta$, $M_c$ and $\mu_s$. Whereas three of them ($v_o$, $M_c$ and $\mu_s$ ) are determined
 through the three minimal conditions
 in terms of the two parameters $\mu_m^2 = 2\mu_f^2$ and $\mu_{inst}$ which are in principle calculable
 from QCD and given in terms of the QCD parameters $g_s(\mu)$ (or $\mu$) and $\Lambda_{QCD}$. The parameter $\beta$ is also
 fixed by the ratio $\beta = \mu_m^2/\mu_f^2 =2 $ based on the fact that the quark condensates are almost flavor
 independent. It has been seen that the four parameters $v_o$, $\beta$, $M_c$ and $\mu_s$ are reduced to two
 independent parameters and well determined from the low energy dynamics of mesons. Of interest, they
 lead to consistent predictions for the light quark masses and pseudoscalar meson masses as well as
 for the lowest nonet scalar meson masses and
 mixing at the leading order. In particular, the resulting quark condensate is consistent with the QCD prediction.
 In general, once the basic parameters at the leading terms are determined, all the higher order corrections
 from momentum and quark mass expansions can be systematically
 calculated based on the present considerations without involving
 any additional parameters. This is because all the couplings of higher order corrections in the
 momentum expansion only depend on the independent parameters $v_o$ (or $\mu_s$) and
 $M_c$. More general description on the effective chiral quantum field
 theory of mesons with including possible high order terms and also the vector and axial vector mesons is beyond
 our present purpose and will be considered elsewhere.

\hspace{3.0cm}

 {\bf  Acknowledgement}

  This work was supported in part by the key projects of Chinese Academy of Sciences
  and National Science Foundation of China.

 \appendix

 \section{ Derivation of the Effective Chiral Lagrangian by New Finite Regularization Method}

  To obtain the effective chiral Lagrangian for mesons given in the text, we need to integrate over the quark fields
  (which is equivalent to calculate the Feynman diagrams of quark loops) from the following chiral Lagrangian
 \begin{eqnarray}
 {\cal L}_{eff}^q & = & \bar{q}\gamma^{\mu}i\partial_{\mu}q +
 \bar{q}_L\gamma_{\mu}{\cal A}_L^{\mu}q_L + \bar{q}_R\gamma_{\mu}{\cal A}_R^{\mu}q_R
 - [\ \bar{q}_L(x)(\Phi(x)-M) q_R(x) + h.c. \ ] \ \nonumber \\
 \end{eqnarray}
 The above Lagrangian is invariant under transformations of the global chiral
 symmetry $U(3)_L \times U(3)_R$  in the limit $m_i \rightarrow 0$ ($i=u,d,s$),
 \begin{eqnarray}
  q_L(x)\equiv P_+ q(x)\rightarrow g_L q_{L}(x), \quad  q_R(x)\equiv P_- q(x)\rightarrow g_R q_{R}(x);
  \quad \Phi(x) \rightarrow g_L \Phi(x) g_{R}^{\dagger}
 \end{eqnarray}
 Using the method of path integral, the effective Lagrangian of mesons is evaluated via
\begin{equation}
\int[d\Phi] exp \{i\int d^{4}x{\cal L}^{M}\}= Z_0^{-1} \int [d\Phi][dq][d\bar{q}] exp \{i\int d^{4}x{\cal
L}_{eff}^q \}.
\end{equation}
The functional integral of right hand side is known as the determination of the Dirac operator
\begin{equation}
 \int[dq][d\bar{q}]
exp \{i\int d^{4}x{\cal L}_{eff}^q \} = \det(i{\cal D}).
\end{equation}
To obtain the effective action, it will be useful to go to Euclidean space via the rule
$\gamma_0 \to i \gamma_4$,
$G_0 \to i G_4$, $x_0 \to -i x_4$ and to define the Hermitian operator
\begin{eqnarray}
S_E^{M} & = & \int d^4x_E{\cal L}^{M}_{E}=\ln \det i{\cal D}_E  \nonumber \\
 & = & \frac{1}{2}[\ln \det i{\cal D}_E + \ln
\det (i{\cal D}_E) ^{\dagger} ] + \frac{1}{2}[\ln \det i{\cal D}_E - \ln \det (i{\cal D}_E)^{\dagger} ] \equiv
S_{ERe}^M + S_{EIm}^M
\end{eqnarray}
with
\begin{eqnarray}
S_{ERe}^{M}=\int d^4x_E {\cal L}_{Re}^M & = & \frac{1}{2}\ln
\det (i{\cal D}_E (i{\cal D}_E)^{\dagger}) - \ln Z_0
\equiv \frac{1}{2}\ln \det\Delta_E - \ln Z_0 \nonumber
\\ S_{EIm}^{M}=\int d^4x_E {\cal L}_{Im}^M & = & \frac{1}{2}\ln \det (i{\cal D}_E /(i{\cal
D}_E)^{\dagger}) \equiv \frac{1}{2}\ln \det\Theta_E
\end{eqnarray}
where the imaginary part ${\cal L}_{Im}^M$ appears as a phase which is related to the anomalous terms and will not
be discussed in the present paper. The operators in the Euclidean space are given by
\begin{eqnarray}
& & i{\cal D}_E = -i\gamma\cdot \partial- \gamma\cdot {\cal A}_L P_L  -\gamma\cdot
{\cal A}_R P_R + \hat{\Phi} P_R +
 \hat{\Phi}^{\dagger} P_L \, \nonumber \\
& & (i{\cal D}_E)^{\dagger} = i\gamma\cdot \partial + \gamma\cdot {\cal A}_R P_L
+ \gamma\cdot {\cal A}_L P_R +
\hat{\Phi}^{\dagger} P_R + \hat{\Phi} P_L
\end{eqnarray}
with $ \hat{\Phi} = \Phi - M$ and $P_{\pm} = (1\pm \gamma_5)/2$. $\Delta_E$ is regarded as a matrix in coordinate
space, internal symmetry space and spin space. Noticing the following identity
\begin{eqnarray}
 \ln \det O = Tr \ln O
\end{eqnarray}
with $Tr$ being understood as the trace defined via
\begin{eqnarray}
 Tr O = tr \int d^4x < x | O | y > |_{x=y}
\end{eqnarray}
Here $tr$ is the trace for the internal symmetry space and $<x | O| y>$ is the coordinate matrix element defined
as
\begin{eqnarray}
 < x | O_{ij} | y > = O_{ij} (x) \delta^4(x-y)\ , \qquad
 \delta^4(x-y)= \int_{-\infty}^{\infty} \frac{d^4k}{(2\pi)^4} e
 ^{i k\cdot (x-y) }
\end{eqnarray}
For the derivative operator, one has in the coordinate space
\begin{eqnarray}
 < x | \partial^{\mu} | y > =
 \delta^4(x-y)( - i k^{\mu} + \partial_y^{\mu})
\end{eqnarray}

With these definitions, the operator $\Delta_E$ in the Euclidean space is given by
\begin{eqnarray}
  < x| \Delta_E | y >  =
 \delta^4(x-y)  \Delta_E^k
\end{eqnarray}
with
\begin{eqnarray}
& & \Delta_E^k = k^2 + \Delta_E  \equiv \Delta_0 + \tilde{\Delta}_E \\
 & & \Delta_0 = k^2 + \bar{M}^2 \\
 & & \tilde{\Delta}_E  = \left(\hat{\Phi}\hat{\Phi}^{\dagger} -
 \bar{M}^2\right)P_R + \left(\hat{\Phi}^{\dagger} \hat{\Phi} -
 \bar{M}^2\right)P_L - i\gamma\cdot D_E \Phi P_L - i\gamma\cdot D_E \Phi^{\dagger}
 P_R \nonumber \\
 & & - \sigma_{\mu\nu} {\cal F}_{R\mu\nu} P_L - \sigma_{\mu\nu} {\cal F}_{L\mu\nu}
 P_R + (iD_{E\mu})(iD_E^{\mu})+  2k\cdot(iD_E)
\end{eqnarray}
with
\begin{eqnarray}
& & iD_E\Phi = i\partial \Phi + {\cal A}_L \Phi - \Phi {\cal A}_R \\
& & iD_E = i\partial + {\cal A}_R P_L  + {\cal A}_L P_R
\end{eqnarray}
Where $\bar{M}$ is the supposed vacuum expectation values (VEVs) of $\hat{\Phi}$, i.e., $<\hat{\Phi}> = \bar{M} =
diag.(\bar{m}_u, \bar{m}_d, \bar{m}_s)$. With this convention, it will be seen that the minimal
conditions of the effective potential are completely determined by the lowest order terms
 up to the dimension four  $\left(\hat{\Phi}\hat{\Phi}^{\dagger} -\bar{M}^2\right)^2$ in
the effective chiral field theory of mesons.
 Regarding $\tilde{\Delta}_E$ as the interaction term and taking
\begin{eqnarray}
 Z_0 = (\det\Delta_0)^{1/2}
\end{eqnarray}
Thus the effective action in the Euclidean space can be written as
\begin{eqnarray}
S_{ERe}^M  =  \frac{1}{2}\ln \det\Delta_E (\Delta_0)^{-1} = \frac{1}{2} Tr \ln \frac{\Delta_E}{\Delta_0}
\end{eqnarray}
Using Schwinger's proper time technique \cite{JS}
\begin{eqnarray}
 \ln \Delta  =
-\int_{\tau_0}^{\infty}\frac{d\tau}{\tau} e^{-\tau \Delta} -\left( \ln \tau_0 + \gamma +
 \sum_{n=1}^{\infty} \frac{(-\tau_0 \Delta)^n}{n\cdot n !} \right)
\end{eqnarray}
 the effective action is found in the coordinate space to be
 \begin{eqnarray}
S_{ERe}^M & = & -{1\over 2} \int d^{4}x_E \frac{d^{4}k}{(2\pi)^{4}} tr\int^{\infty} _{0}{d\tau\over \tau} \left(
e^{-\tau \Delta_E^k } -e^{-\tau\Delta_{0}} \right)
\nonumber \\
 & = & -{1\over 2} \int
d^{4}x_E\frac{d^{4}k}{(2\pi)^{4}} tr\int^{\infty} _{0}{d\tau\over \tau} e^{-\tau\Delta_{0}} \left( e^{-\tau
\tilde{\Delta}_E^k -\frac{\tau^2}{2} [\bar{M}^2 \ \tilde{\Delta}_E^k] } - 1 \right)
\end{eqnarray}
where we have used the identity $e^A e^B = e^{A + B + \frac{1}{2}[ A \ B]}$. Treating $\tilde{\Delta}_E$ as
perturbative interaction and making an expansion, we have

 \begin{eqnarray}
S_{ERe}^M
 & = & -{1\over 2} \int
d^{4}x_E\frac{d^{4}k}{(2\pi)^{4}} tr \int^{\infty} _{0}{d\tau\over \tau} e^{-\tau\Delta_{0}} \sum_{n=1}^{\infty}
\frac{(-1)^n}{n!} \tau^n  [\tilde{\Delta}_E^k -\frac{\tau}{2} [\bar{M}^2 \ \tilde{\Delta}_E^k]^n
 \end{eqnarray}
For the given order of expansion, the integral over $\tau$ can be performed by using the following integral
 \begin{eqnarray}
 \int^{\infty} _{0}{d\tau\over
\tau} e^{-\tau\Delta_{0}} \tau^n = (n-1)! \Delta_0^{-n}
\end{eqnarray}

For the integral over momentum, which involves quadratically and logarithmic divergence. In order to maintain the
gauge invariance and meanwhile keep the quadratic term, the new finite regularization method proposed recently in
ref.\cite{YLW} should be adopted for the momentum integral.
\begin{eqnarray}
& & I_2 = \int \frac{d^4k}{(2\pi)^4} (k^2 + \bar{M}^2)^{-1} \to
I_2^R = \frac{1}{(4\pi)^2} T_2 (\frac{\mu^2}{M_c^2}) \\
 & & I_0 = \int
\frac{d^4k}{(2\pi)^4} (k^2 + \bar{M}^2)^{-2} \to I_0^R = \frac{1}{(4\pi)^2} T_0 (\frac{\mu^2}{M_c^2})
\end{eqnarray}
with the consistent conditions\cite{YLW}
\begin{eqnarray}
I_{2\mu\nu}^R  = \frac{1}{2} g_{\mu\nu} I_2^R ,\qquad I_{0\mu\nu}^R  = \frac{1}{4} g_{\mu\nu} I_0^R
\end{eqnarray}
where $L_2$ and $L_0$ are given in text. The superscript `R' means regularized one.

 With these analyzes, the effective chiral Lagrangian can be systematically obtained in the expansion
 of momentum and current quark masses by transforming back to the Minkowski spacetime.
 In the text, we only kept those terms needed for our purpose in the present paper.


\begin{thebibliography}{99}
 \bibitem{CM} See e.g.: J.J. Sakurai, {\it Currents and Mesons} (University of Chicago Press, Chicago, 1969).
  \bibitem{PCAC} M. Gell-Mann and M. Levy, Nuovo Cimento, {\bf 16} (705) 1960; \\
  K.C. Chou, Soviet Physics JETO {\bf 12} 492 (1961).
  \bibitem{NJL} Y. Nambu and G. Jona-Lasinio, Phys. Rev. {\bf 122} 345 (1961).
  \bibitem{INST} G. 't Hooft, Phys. Rev. Lett. {\bf 37} 8 (1976), Phys. Rev. {\bf D14} 3432 (1976);
  Err. Phys. Rev. {\bf D18} (1978) 2199.
  \bibitem{LSM} J. Schwinger, Ann. Phys. {\bf 2} 407 (1957);
  M. Gell-Mann and M. Levy, Nuovo Cimento, {\bf 16} (705) 1960; J. Schechter and Y. Ueda, Phys. Rev. {\bf D3} 2874
  (1971).
  \bibitem{PDG} D.E. Groom et al. [Particle Data Group Collaboration], Eur. Phys. J. {\bf C15} 1 (2000).
  \bibitem{E7912} E. M. Aitala et al. (E791 collaboration), Phys. Rev. Lett. {\bf 89} 121801 (2002).
   \bibitem{E7911} E. M. Aitala et al. (E791 collaboration), Phys. Rev. Lett. {\bf 86} 765 (2001).
   \bibitem{sigma} See e.g. Conference on ``Possible existence of the light $\sigma$ resonance and its
   implications to hadron physics'', Kyoto, Japan 11-14th June 2000, KEK-proceedings/2000-4 .
   \bibitem{YLW} Y. L. Wu, hep-th/0209021, Int. J. Mod. Phys. A 18  (2003).
   \bibitem{INST2} G. 't Hooft, hep-ph/9903189, 1999.
    \bibitem{JS} J. Schwinger, Phys. Rev. {\bf 93} (1954) 613.
 \bibitem{kappa} R.J. Jaffe, Phys. Rev. {\bf D15} 267, 281 (1977); \\
    M.D. Scadron, Phys. Rev. {\bf D26} 239 (1982); \\
    E. Van Beveren, T. A. Rijken, K. Metzger, C. Dullemond, G. Rupp and J. E. Ribeiro, Z. Phys. {\bf C30} 615
    (1986); \\
    S. Ishida, M. Ishida, T. Ishida, K. Takamatsu and T. Tsuru, Prog. Theor. Phys. {\bf 98} 621 (1997); \\
    D. Black, A. H. Fariborz, F. Sannini and J. Schechter, Phys. Rev. {\bf D58} 054012 (1998); \\
    J. A. Oller and E. Oset, Phys. Rev. {\bf D60} 074023 (1999); \\
    M. Napsuciale, hep-ph/0204170, 2002; \\
    N. A., T\"{o}rnqvist, hep-ph/0204215, 2002.
  %  \bibitem{DaiWu} Y.B. Dai and Y. L. Wu, in preparation.
 \end{thebibliography}
\end{document}